# A first principles analysis of the Colossal Ion Conductivity cubic Zirconia structural hypothesis and disorder, mechanics, and space charge mechanistic hypotheses

M.F. Francis
Los Alamos National Laboratories, Los Alamos, New Mexico, 87545

**Abstract**

Colossal Ion Conductivity (CIC) is the phenomenon in which large changes in oxygen conductivity are observed when a solid state oxygen electrolyte is placed in a heterolayer super lattice. Several hypotheses have been posed concerning both structural changes of the ion conducting material and of the ion conducting mechanism. The posed structural hypotheses are oxygen sub lattice melting, phase change of the ion conducting material, and that the heterolayer environment does not induce a phase change of the ion conducting material. The posed mechanistic hypotheses are disorder, mechanics, and space charge. Here, first principles molecular dynamics and statics of the ion conducting interface are performed providing evidence that under CIC conditions no sub lattice melting nor phase change has occurred and that the electrolyte remains unchanged. A discussion of the posed mechanistic hypotheses is given and cast as a series of mathematical statements. When these hypotheses are compared with first principles calculations and literature results, they fail to capture CIC; a new alternative dipole hypothesis is offered, which also fails to capture CIC; further analysis of the literature points to CIC being an interface mediated phenomenon, similar to grain boundary diffusion, where the *operando* mechanism remain unclear.

## Introduction

One of the great challenges of our era is leaving behind a dependence on fossil fuels and building an environmentally benign energy infrastructure. One solution that has been proposed has been the use of fuel cells, capable of extracting energy from fuels at very high efficiencies[1]. The fuel cell is an electrochemical cell consisting of two electrodes separated by an electrolyte[2]. The fuel cell current is the result of the flow of ions from one electrode, through the electrolyte, to the other electrode; and the fuel cell voltage is the result of the ion diffusion energy and the energy of the chemical reactions at the electrodes[2]. Capturing most of the reaction energy in the form of a current avoids several processes that otherwise would create losses in combustion[1,3,4]. This higher energy efficiency makes fuel cells a desirable replacement for combustion technology.

The most commonly used and discussed fuel cell is based on hydrogen[5]. In hydrogen based fuel cells, hydrogen diffuses across the electrolyte as protons, and then reacts with molecular oxygen to create water[2]. The current density is linked to the proton diffusion rate; and the voltage to the hydrogen-oxygen reaction energy per unit proton charge[2]. The reaction between hydrogen and oxygen is facilitated by rare and expensive catalytic materials[6] and requires that only hydrogen is used as a fuel. The limitation to hydrogen requires shipping, as existing oil and gas pipelines explode when exposed to hydrogen gas[7]. The restriction to a single hydrogen fuel may be the main factor limiting the widespread adoption of fuel cells.

An alternative fuel cell is based on oxygen diffusion, utilizes an arbitrary fuel, and relies on abundant materials – solid oxide fuel cells, or SOFCs. In SOFCs, oxygen is dissociated at an electrode and diffuses across the electrolyte in the form of charged oxygen vacancies, and then reacts with a fuel at the opposite electrode[8,9]. The current density is controlled by the oxygen vacancy diffusion rate; and the voltage is controlled by the fuel-oxygen reaction energy per unit oxygen vacancy charge[9,10]. The use of SOFCs allow the utilization of arbitrary fuels, and therefore direct deployment into existing energy infrastructure. Despite the significant SOFC advantages, use has been limited by the slow diffusion of oxygen through the electrolytes and consequential low current densities. This slow oxygen diffusion makes the operating temperature 800 to 1000 °C, when ambient conditions are desired[11-13]. If a scientific solution could be found to facilitate oxygen diffusion, low cost, energy efficient fuel cells, that can utilize an arbitrary fuel could be made a reality.

The prospect of precisely such a solution presented itself in a phenomenon that came to be known as Colossal Ion Conductivity (CIC)[14]. An oxide, A, is grown on top of a layer of another material, B, forming a heterolayer; this A-B- heterolayer was deposited repeatedly, forming a bulk structure, and resulting in a structure known as a heterolayer super lattice. When a yttria stabilized zirconia (A=YSZ), strontium titanate (B=STO), A-B- heterolayer superlattice was grown, diffusion was observed to change by up to eight order of magnitude. The observed changes in diffusion were enough to bring SOFCs into the performance space required for practical application, suggesting SOFCs could replace combustion.

The size of the effect, the potential importance of its impact, and the complexity of the experiment quickly led to debate[15]. Questions related to electron conduction, electrode contamination, interfacial space charge effects, and oxide defects were all posed[15]. Ultimately, further experiments successfully reproduced large changes in diffusion[16-18], as did detailed quantum mechanical simulations[16,19]. CIC is real, but there remains no mechanistic picture describing the origin of the effect, no understanding of the conducting phase, nor a roadmap to engineering future materials. A fundamental, mechanistic understanding of CIC, wherein it is described by a predictive equation, and not observations, would enable materials optimization to maximize this effect in SOFC electrolytes.

The goal of this work is to begin the development of the design enabling, mechanistic understanding that is achievable through an analytical expression. The three major structural and mechanistic hypotheses found the in the literature[16,17,20-29] are converted into mathematical form and tested. This method has the merit of defining what is controlling – interesting, though not causal, phenomenon may be observed, confusing the goal of design. Here, it argued that the conducting phase is the native cubic zirconia, the likely conducting phase is the interface itself, not the near interface bulk, and that all proposed bulk mediated mechanisms fail to predict CIC.

## Method

Properties were evaluated with first principles density functional theory (DFT) as implement in VASP[30-32]. The Perdew-Burke-Ernzerhof (PBE) exchange and correlation functional were used for all simulations[31,32]. This choice of functional was made to be consistent with other first principles studies of CIC[16,19,20]. Careful comparisons of functionals predictions of phenonoma shows no significant change between functionals, making the consistency of functional more importance for facilitating comparison rather than correctly reproducing the desired physics[33].

**Ab initio molecular dynamics**

The highly complex nature of the interface requires the use of *ab initio* molecule dynamics (AIMD). *Ab initio* (AI) methods allow the modeling of complex structures with quantum mechanical accuracy, regardless of the complexity of the chemical and electronic interactions. Molecular dynamics (MD) propagates the system according to *in situ* forces, indispensable in systems where the controlling dynamics and physics are unknown. AIMD allows the hypothesis testing of the CIC interface under operating conditions. With a model interface assembled, molecular dynamics may be used to propagate the structure to obtain temperature dependent structure and dynamics.

The original[14], experimentally observed CIC interface was of 8% YSZ and STO where the YSZ and STO were rotated by 45° with respect to one another and oriented with faces (100)cZrO$_2$/(100)STO. Here, and in the prior *ab initio* studies, the 45° rotated interface was assembled (100)cZrO$_2$/(100)STO, wherein the (100)cZrO$_2$/(100)STO is assumed to be Ti-terminated. The first follow-up computational study directly examined the interfacial environment, with 1 nm of YSZ (5 Zr layers) and 1 nm of STO (4 layers of Sr, 5 of Ti); by comparing oxygen diffusion at the temperatures of 1000, 1500, 2000, 2250, and 2500 K determined a barrier to diffusion of $E_{AIMD}^{diff} = 0.4 \pm 0.1$ eV; this first study significantly verified the observation of CIC in a period where the community was still in debate as to whether the effect was real[16]. Follow-up studies showed used precisely the same structure to show that at 2000 K the oxygen sub lattice melts[20]. The same atomic structures were again used at a later stage, but varying the STO to Al$_2$O$_3$ and KTaO$_3$, verifying that CIC could be mediated by different interfaces[19].

Here, the model CIC interface was composed of 3-layers of STO and 5-layers of cZrO$_2$, rotated 45° with respect to one another and oriented with faces (100)cZrO$_2$/(100)STO. This is different from the previous computational studies in two ways – the STO layer is one atomic layer thinner, and in constructing the (100)cZrO$_2$/(100)STO interlayer two Ti atoms were removed to maintain charge neutrality (Fig 1). These differences are not expected to affect the comparability of the results as results being examined are the dynamics and structure of the electrolyte in the environment, not the interlayer.

The structure of the model supercell is shown in Fig 1 and contains 20 Zr atoms, 12 Sr atoms, 12 Ti atoms, and 76 oxygen atoms, from which 2 Zr were replaced with Y and one oxygen was removed to create an interface containing 10% yttrium stabilized zirconia (YSZ). The structure of cubic zirconia (cZrO$_2$) and strontium titanate (STO) was first determined by minimizing the stress tensor. These stress tensor minimizations were performed with 40 subdivisions of reciprocal space, shifted from the gamma center by 0.5, 0.5, 0.5, resulting in a 10x10x10 mesh for both cZrO$_2$ and STO. The cZrO$_2$ and STO structures were used as input into a model of the CIC interface. The yttrium atom substitution was performed in the second and fourth YSZ planes, with the position selected to maximize the initial distance between the same charged Y. The interfacial model was then subjected to stress tensor minimization in order to remove any artifacts from the addition of the yttrium and compositional vacancies. The resulting interfacial model structure is shown in Fig 1.

During the AIMD convergence criteria of $10^{-4}$ eV for electronic structure, and 0.02 eV/Å for ions, with an automatically determined k-mesh of 2x5x5 was used to propagate the system according to Langevin dynamics[34]. The molecular dynamics was run for a total of 5 ps, at time steps of 1 fs, at a temperature of 1000 K. The original, experimentally observed CIC measurements were between 300 and 673 K placing these AIMD temperatures at slightly above experimental conditions. Within the AIMD community it is common practice to perform simulations at elevated temperatures, here 1000 K compared to the experimental 300-673 K. This elevated AIMD temperature allows the more rapid observation of events, important due to the significant AIMD computational expense. The practice of utilizing an elevated temperature is valid so long as modifying the temperature does not change the underlying phenomenology and allows the observed phenomenology to be extrapolated to desired temperatures. Here, no extrapolation is performed as the main use of AIMD here is to characterize the physical structure of the interface. The AIMD calculations are used to assess melting.

### *Static Properties*

First principles methods were used to determine the intrinsic properties of the systems, and put into physics based models of CIC. The minimum energy pathways and transition states were determined using the nudged elastic band (NEB) technique[35,36]. Stress states were determined through a Hellmann-Feynman analysis[37-40]. The elastic moduli of cubic zirconia were determined according to $c_{11} = \partial\sigma_{11}/\partial\varepsilon_{11}$, $c_{12} = \partial\sigma_{22}/\partial\varepsilon_{11}$, $c_{44} = \partial\sigma_{12}/\partial\varepsilon_{12}$, and $B = c_{11} + 2c_{12}$. The resulting cZrO$_2$ elastic properties are shown in Table 1. The elastic dipole[41], $\boldsymbol{P}$, is given as the negative of the product of the cell volume, $V$, and the homogenous cell stress, $\boldsymbol{\sigma}$, according to $\boldsymbol{P} = -\boldsymbol{\sigma}V$. It has been shown that errors in the elastic dipole of charged systems cancel upon differencing[42], making the activation dipole free of charge related errors, and directly applicable for discussion of CIC mechanism. The activation volume, $V^{diff}$, was determined by taking the hydrostatic components of the homogeneous stresses at the transition state, $\sigma_h$, dividing by the bulk modulus, and multiplying by the cell volume $V^{diff} = \sigma_h V/B$. The statics calculations are used to assess disorder, mechanics, and the space charge hypothesis.

### Results

**Structural hypotheses**

It has been posited that the heterolayer environment, (I a), results in the melting of the oxygen sublattice[16,20], while the cations remain crystalline, providing liquid-like oxygen conductivity, (I b), induces a YSZ phase change[21,22], or, (I c), allows the YSZ to remain in its native cubic phase[23].

Each of the phase hypotheses may be addressed by using a pair correlation function[43], $p_{ij}(r)$. The pair correlation function may be determined by taking the average number of $ij$ pairs, $n_{ij}(r, \Delta r)$, a distance $r$ away from each other, within a cross section of $\Delta r$, and dividing by the total number of pairs, $N_{ij}$, giving $p_{ij}(r, \Delta r) = n_{ij}(r, \Delta r)/N_{ij}$. Liquid, fluid, and amorphous solid pair correlation functions demonstrate a single peak at the spacing of a single atomic separation, associated with nearest neighbor clustering, and a smooth structure for higher atomic spacing, associated with the structurally disordered bulk. Crystalline solid pair correlation functions demonstrate well-ordered peaks at characteristic positions. Examination of the pair correlation function of $ZrO_2$ in the CIC model can be used to determine the phase of the oxygen sub lattice and the ion conducting material. Figure 2 shows the pair correlations of the zirconia where the black data represents $cZrO_2$ and the blue data represents the AIMD *in situ* sampled zirconia.

**Sub lattice melting (I a)**. Shown in Fig 2a is the OO pair correlation function in the (100) ion conducting plane. The pair correlation function shown in Fig 2a was created from an analysis of structures generated every 1 fs over the course of the last 1.4 ps of 5 ps of simulation. This AIMD simulation time of 5 ps compares to prior simulation times of 6 and 7 ps[16,19,20]. Fig 2a has nonzero values at zero atomic separation associated with OO pairs occupying the same positions in parallel planes and three defined peaks at 2.75, 3.85, and 5.5 Å. Fig 2a shows multiple peaks at different positions indicating that the oxygen sub lattice has not melted and remains in a well ordered phase. AIMD simulations at higher temperatures have shown sub lattice melting[20], and it is true that the sub lattice melts, but the evidence is that at CIC relevant conditions the sub lattice remains unmelted. The time scale required to observe oxygen sublattice melting is that of the oxygen diffusive processes making the simulation lengths sufficient.

**Phase change (I b,c)**. Shown in Figs 2b,c are the ZrZr and ZrO pair correlation functions in the (100) ion conducting plane. Shown in blue are pair correlation functions at 1000 K and shown in black are pair correlation functions from cubic $ZrO_2$ at 0 K in the interfacial environment. The black zero data shows dispersion in the pair correlation functions; this dispersion is not due to any temperature but to structural changes induced by the interface. The ZrZr correlation function shows a peak about zero, again associated with Zr pairs occupying the same position in parallel plans, and three subsequent peaks at 2.65, 3.65, and 5.5 Å. The ZrO correlation function shows no peak about zero, indicating Zr and O atoms in parallel (100) planes do not occupy the same position, and shows defined structure with peaks at 1.85, 4.15, 5.65, and 6.55 Å. Because $ZrO_2$ is binary, precisely two correlation functions are required to characterize the phase, and when it is revealed that the cubic $ZrO_2$ and 1000 K structures have the same ZrZr and $ZrO_2$ correlation functions it is revealed that CIC $ZrO_2$ has not undergone a phase change and remains cubic. The time scale required to observe phase change is tied to the phase change mechanism, which is unknowable without knowing the final phase, this makes discussion of whether the time scale is long enough a matter of speculation. Though the evidence against a phase change must be tempered by the time scale, where a more detailed discussion is provided in the supplementary information (SI), CIC can and has been observed in $cZrO_2$ structures[19] serving as strong, positive evidence of the *operando* phase as $cZrO_2$.

**Mechanistic hypotheses**

The prevalent mechanistic hypotheses are, (II a), a heterolayer induced disorder[16,24], (II b and c), a mechanical effect[17,25-27], and/or (II d), a space charge mediated interfacial effect[28,29]. Diffusion, $D$, may be written as a function of a dimensional constant, $n$, an attempt frequency, $v$, carrier concentration, $c$, activation entropy, $\Delta S^{act}$, and activation energy, $\Delta E^{diff}$, according to

$$D = nvce^{\Delta S^{act}/k_B}e^{-\Delta E^{diff}/k_BT}.$$

**Disorder (II a)**. The structures represented in Figs 1,2 are observably disordered and have lead to various researchers proposing that disorder is responsible for CIC. Disorder may affect the diffusion rate by modifying the entropy of the ground and transition states creating a change in activation entropy, $\Delta S^{act}$.

CIC changes have demonstrated a marked Arrhenius temperature dependence, with that temperature dependency changing from interface to interface[17,25-28,44-47]. One example of the Arrhenius dependence is that of Feng Li et al (Fig 3)[19]. Feng Li et al have performed AIMD simulations with YSZ, varying the straining material. Feng Li et al observed that they were able to mediate the CIC effect by changing the interface, but that in each case the change showed up as an activation barrier effect, and not an entropy effect. CIC is controlled by enthalpy, not entropy, and that while disorder exists[48], it is not predictive of the novel CIC properties.

**Mechanics (II b)**. The correlation of the CIC effect with the structural misfit, as in Fig 3, has lead to several researchers proposing a mechanical effect. Underlying the mechanics hypothesis is the belief that with each hop, a diffusion volume, $V^{diff}$, must be displaced. When a stress is applied to the solid, $\sigma_h$, it couples to this diffusion volume forming a mechanical work term[25], given as

$$\Delta E_h^{diff} = \sigma_h V^{diff}.$$

The activation energy and diffusion volume for the three oxygen vacancy diffusion pathways has been calculated and is shown in Table 2. The activation energy is shown for the charge neutral and +2 charged vacancy. The uncharged oxygen vacancy has activation energies of 2.52, 2.63, and 10.74 eV where the charged oxygen vacancy shows an activation barrier of 1.165 eV. The experimentally measured bulk oxygen vacancy diffusion barrier is 1.1±0.1 eV [49], emphasizing that the vacancy is charged, and verifying the correct oxygen migration pathway.

The comparison of the DFT and experimental activation energies clearly identifies the diffusion pathway, but also identifies the mechanics hypothesis to be incomplete. CIC observations are of a change in activation energy of ~0.4 eV[49]; if a 6 % epitaxial strain is assumed, with a -4 % out of plane compensation. The elastic moduli of cubic Zirconia predict a hydrostatic stress of ~22 GPa. When coupled to the formation volume the resulting change in activation barrier is the small value of 0.00065 eV. Mechanical effects are too small to account for the ~0.4 eV change in activation barrier.

**An alternative mechanical hypothesis, dipole theory (II c)**. Though the mechanical hypothesis fails to predict CIC effects, CIC demonstrates texture and orientation effects. A texture effect is one in which one, or both, of the materials used to form the A/B heterolayer super lattice are rotated with respect to the other, resulting in a change of crystal face, or texture, at the interface, and a consequential change in CIC. Texture effects have been observed when comparing varying YSZ[49] and SDC[44] textures. Modifying the texture of the interface rotates the migration path with respect to the environment.

An alternative mechanical hypothesis may be offered from dipole theory, which allows a mechanical description of this orientation effect. It is imagined that the hopping process begins at a single ground state, Fig 4a, passes through a transition state, Fig 4b, and arrives at another ground state, Fig 4c. Fig 4 describes what is imagined by this hopping process, where Zr is showng in green, and oxygen in red; the dashed black circle indicates the hopping oxygen atom, and the dashed grey circle indicates the hopping vacancy; the black arrows indicate the direction of the displacements of the oxygen atoms. If the ground states have a symmetric structure, they will generate a spherical stress field in the solid, Figs 4a and 4c, and if the transition state passes through a constrained space, it will create an asymmetric stress, having a different in plane than out of plane structure.

The dipole model allows a directionality to be captured, that may have been lost by the mechanical picture, and increase the overall response to the epitaxial strain field. The change in energy due to an applied strain, $\varepsilon$, may be written as a dot product with an activation dipole[41], $\Delta \boldsymbol{P}^{act}$. That activation dipole is given as the difference between the ground state dipole, $\boldsymbol{P}^G$, and one for the transition state dipole, $\boldsymbol{P}^{TS}$, according to $\Delta \boldsymbol{P}^{act} = \boldsymbol{P}^{TS} - \boldsymbol{P}^G$, and gives a change in migration barrier

$$\Delta E_{dip}^{mig} = \Delta \boldsymbol{P}^{act} \cdot \boldsymbol{\varepsilon}.$$

The activation dipole for the three oxygen vacancy diffusion pathways has been calculated and is shown in Table 2. In each case, the activation dipole is asymmetric, as predicted by the conceptual discussion of the activation and ground state, Fig 4. The asymmetry of the activation dipole may account for the experimental texture effects, but when again the magnitude of the effect is examined it is found again to be too low. If the 6 % epitaxial strain is assumed, with the -4 % out of plane compensation, the resulting change in energy is ~0.005 eV. Dipole theory accounting for the orientation effects has increased the mechanical effect by an order of magnitude, but it remains two orders of magnitude too small.

**Space charge (II d)**. Underlying the space charge hypothesis is the belief that charge will agglomerate at the interface, change the local potential, and consequentially the diffusion barrier. The change in potential, $\Delta\varphi$, mediates a change in diffusion barrier, $\Delta E^{diff}$, through an electrostatic effect when the charge of the transition state, $q_{TS}$, and that of the ground state, $q_R$, are different according to

$$\Delta E_{pot}^{diff} = q_{TS}\Delta\varphi_{TS} - q_R\Delta\varphi_R.$$

The influence of the space charge on the activation barrier may be tested by directly examining the activation barrier in the interfacial structure (Fig 1). By examining the activation inside of the interface, any band bending that and consequential potential shift will be naturally represented by DFT. The interfacial structure has six planes of oxygen for possible migration. Two of these planes of oxygen are those forming the boundary between the STO/YSZ and YSZ/STO interfaces. The remaining four are shown in Fig 1 and are labelled 1, 2, 3, and 4. The 1 and 3 planes are structurally symmetric, as are the 2 and 4 planes. The activation barrier to diffusion across the [100] planes for the 1 & 3 and 2 & 4 planes has been calculated shown in Table 3.

These activation barriers are calculated using NEB and used a Y free interface, with the appropriate charge compensation, to uniquely identify the charge and band bending effect. The data shown has two values for each entry. This is the result of the intrinsic structural distortions changing the value of the forward and backward reactions, resulting in a fluctuation effect. In each case, the direct calculation of activation barrier in the interfacial structure showed an increase in the activation barrier, not a decrease.

The direct calculation of the change in activation energy in the interface environment gives a few insights about the mechanistic origins of CIC. It shows that band bending and potential shifts are not responsible for CIC; the interface environment directly creates the band banding. Furthermore, if there had been some nonlinearity in the elastic dipole, resulting in a larger effect with larger strains, that nonlinearity would have shown up in the interface environment. This result reinforces the refuting of the hydrostatic and dipole mechanical hypotheses.

## Discussion

**Structural hypotheses, an alternative**.

**The interface**. The three posed structural hypothesis have each implicitly or explicitly assumed that the CIC conducting structure is the bulk electrolyte. When the structural hypotheses are tested, they point in the direction of a $cZrO_2$ phase, while every major mechanistic hypothesis fails. Despite this apparent mismatch, AIMD simulations of this $cZrO_2$ structure reproduce CIC rates of diffusion[16,19,20]. One possibility of the apparent mismatch between expectation and measurement is the incorrectness of the expectation. It could very well be that it is not the electrolyte, but the interface between the electrolyte and the straining material which is the *operando material*. If the interface is the conducting structure, one would expect that interfacial conductivity had been separately measured and found to be much lower than that of the bulk, and this is exactly the case. Bulk and interfacial diffusion barriers have been measured and separated for zirconia/alumina and it shown that the interface activation barrier is 0.5 eV lower than the bulk[50,51]. For zirconia/yttria, the barrier for the interface is 0.13 eV lower than the bulk. In zirconia/magnesia again, the interface has a barrier of 0.5 eV lower than the bulk[50,51]. These literature findings, combined with the failings of the major bulk mediated hypotheses point strongly to the interface as the *operando* material.

## Conclusions

Here, the major structural and mechanistic hypotheses concerning Colossal Ion Conductivity (CIC) have been scrutinized. A summary of the hypotheses, the required physics, and the evidence for and against is given in Table 4, where less frequently mentioned hypotheses are given in the SI. The net conclusion is that the phase of the conducting material likely remains a cubic zirconia phase, but that each of the bulk mechanistic hypotheses offered fail to capture observed CIC, and that the effect is likely an interfacial phenomenon. This interface mediated phenomenon appears to be controlled by elastically mediated internal energy changes, meaning likely on of the major hypotheses, but active in the interface instead of bulk, and that it will ultimately be the result of a set of interfacial or materials properties that are designable. In the body of this manuscript is described the physics of each hypothesis that has been offered to date. These mathematical and physics castings of hypotheses are readily adaptable to the careful design experiments or simulations. Though the design of CIC is out of reach for now, there is good reason to believe that the community at large will be able to solve this problem, and that facile design of solid state ion conducting materials will soon be a reality.

**Competing financial interests.** The author declares no competing financial interests.

**Author Contributions.** This work was performed solely by MFF.

**Acknowledgements.** This work was performed at Los Alamos National Laboratory, operated by Los Alamos National Security Administration, LLC, for the National Nuclear Security Administration of the US Department of Energy under contract DE-AC52-06NA25396. Support for this work was provided by the Director's Office under the Director's Fellowship program.

**Materials and Correspondence.** Los Alamos National Laboratories, Los Alamos, New Mexico, 87545. mff7d@virginia.edu.

**Figures**

**Fig 1**. Cross sectional view of the ground state structure of the model CIC interface. The structure is modeled after the experimental interface which first showed CIC. The (100)cZrO$_2$/(100)STO oriented 45° with respect to one another and where the 10% of the Zr have been replaced with Y, forming 10% YSZ.

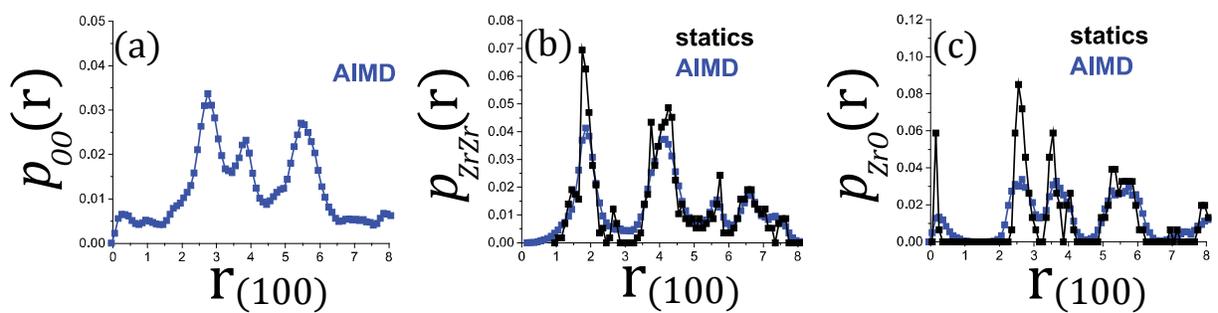

**Fig 2**. Pair correlation functions of the, (a), O-O, (b) Zr-Zr, and (c), Zr-O atomic distances in the model CIC interface. Zero temperature data shown in black and 1000 K molecular dynamics data shown in blue.

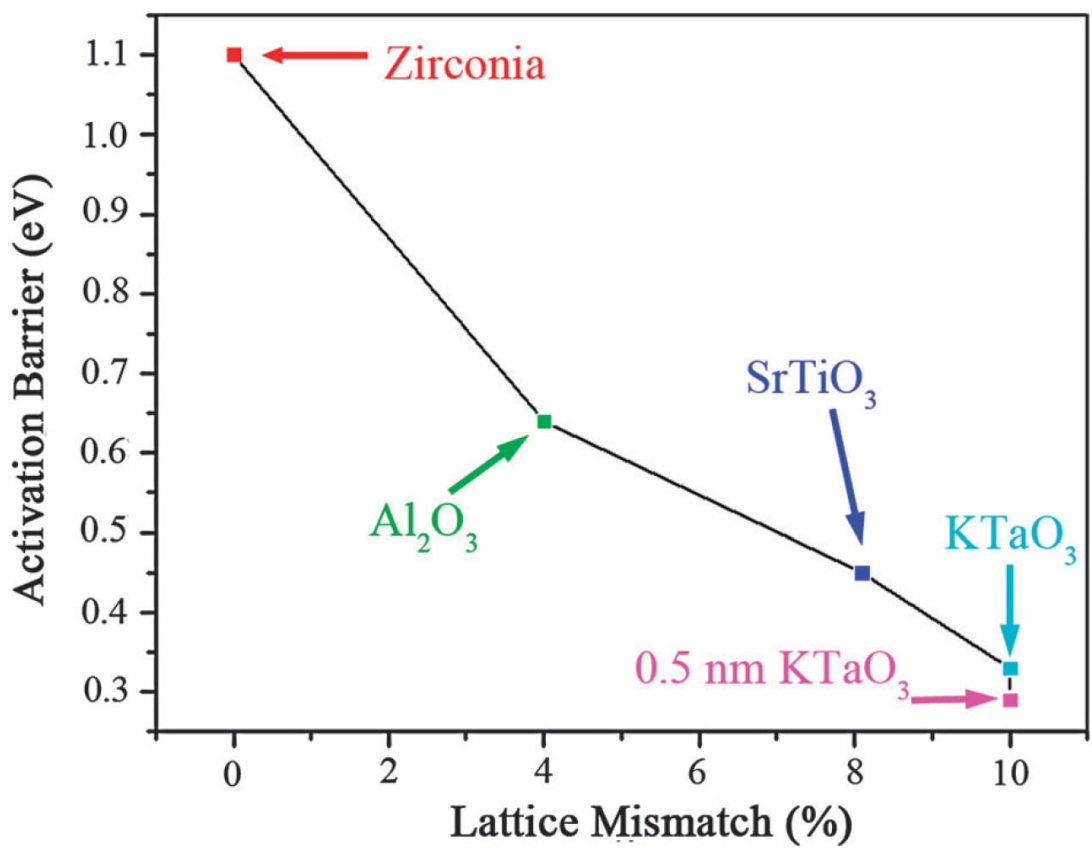

**Fig 3**. AIMD of CIC in YSZ as a function lattice mismatch for various straining materials by Feng Li *et al*[19]. The activation barrier to diffusion was determined from the change in diffusion with temperature according to $E^{diff} = -\partial ln[D]/\partial ln[1/k_B T]$. The diffusion was determined from the mean square displacement, $\langle r^2 \rangle$, as a function of time, $t$, according to the Einstein relation $\langle r^2 \rangle = 6Dt$. The temperature was varied as T = 1000, 1500, 2000, 2250, and 2500 K where each simulation was carried out for a total of 7 ps. Unstrained, bulk cZrO₂ is shown in red; the unstrained, bulk cZrO₂ is examined as a control. The influence of the straining material was examined by taking the same structure as in Fig 1, but varying the straining electrolyte; Al₂O₃ cZrO₂ in green, SrTiO₃ strained cZrO₂ in purple, KTaO₃ strained cZrO2 in light blue. The possible role of thickness of the electrolyte was examined with a 0.5 nm YSZ sample, but the same amount of KTaO₃ as in pink.

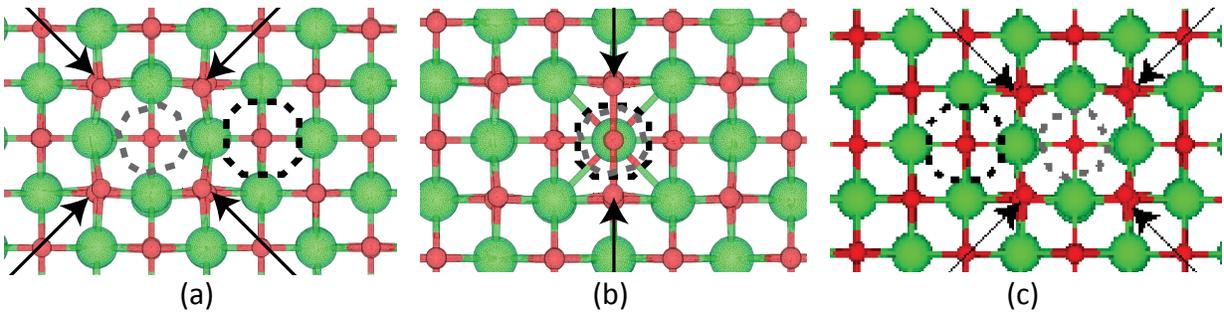

**Fig 4**. Schematic of vacancy mediated oxygen atom hopping in cubic $ZrO_2$. Shown is the, (a), initial, (b), transition, and (c), final state of the oxygen hopping process. Shown in green are the Zr atoms and in red the oxygen atoms. The diffusing oxygen atom is highlighted by the presence of the dashed black circle. The lattice that the oxygen species jumps into in (a) and out of in (c) is visibly spherically distorted and the transition state shown in (b) asymmetrically distorted.

**Tables**

| | |
|---|---|
| $c_{11}$ | 554.88 ± 7.30 GPa |
| $c_{12}$ | 130.66 ± 2.58 GPa |
| $c_{44}$ | 213.92 ± 0.89 GPa |
| $B = c_{11} + 2c_{12}$ | 816.20 ± 8.16 GPa |

**Table 1**. Elastic Properties of cubic ZrO$_2$. These elastic properties have been calculated from DFT according to the protocol described in the methods section.

| Path | Charge (electrons) | Activation Energy (eV) | Activation Volume (Å$^3$) | Activation Dipole (eV) |
|---|---|---|---|---|
| tet to oct | 0 | 2.631 | -0.00430 | -0.0377, -0.0388, -0.0377, 0.0000, 0.0000, 0.0902 |
| [101] tet to tet | 0 | 10.74 | -0.00465 | -0.0407, -0.0353, -0.0407, 0.0000, 0.0000, 0.0821 |
| [010] tet to tet | 0 | 2.516 | 0.01579 | -0.0407, -0.0353, -0.0407, 0.0000, 0.0000, 0.0821 |
| tet to oct | +2 | 1.165 | -0.00475 | -0.0416, 0.0048, -0.0416, -0.0037, -0.0037, 0.0424 |
| [101] tet to tet | +2 | 11.616 | -0.00516 | 0.0880, 0.1118, -0.0676, -0.1294, 0.0350, 0.0492 |
| [010] tet to tet | +2 | 1.165 | 0.00772 | -0.0452, -0.0039, -0.0452, 0.0000, 0.0000, 0.0396 |

**Table 2**. Activation properties of the mobile vacancy state in cubic ZrO$_2$. The first column describes the path; "tet to oct" refers to the oxygen vacancy hopping from the tetrahedral position to the octahedral position, and then back to the tetrahedral, within with fcc-Zr sub lattice; "[101] tet to tet" refers to the hopping from one tethedral position to the other and along the [101] direction; "[010] tet to tet" refers to the hopping from one tethedral position to the other and along the [010] direction. The second column refers to the charge state in units of electrons. The third column is the activation energy in units of eV. The fourth column is the activation dipole given in the form $\Delta \boldsymbol{P}^{act} = (P_{11}, P_{22}, P_{33}, P_{12}, P_{23}, P_{13},)$ in units of eV.

| plane | $\Delta E^{mig}$ (eV) |
|---|---|
| 1, 3 | 1.41, 2.53 |
| 2, 4 | 2.74, 1.61 |

**Table 3**. Activation barrier to migration of the oxygen vacancy in the model CIC interface environment. The model interface environment is that shown in Fig 1, where for this test the Y have been replaced with Zr and electrons exchanged to maintain the appropriate electrostatic environment.

| I. Structural hypotheses | Required Proof | Evidence for, against, and conclusion. |
|---|---|---|
| a. No phase change | For all $i, j$, $p_{ij}^{YSZ}(r) = p_{ij}^{CIC}(r)$. | For, no change has been observed in $p_{ij}^{CIC}(r)$ under CIC operating conditions (Figs 2). CIC has been observed in AIMDs of the unchanged cZrO$_2$ environment[17,19]. Against, the simulation is not long enough to observe the required transformation. |
| b. Sublattice melting | $p_{OO}^{CIC}(r)$ shows no defined peaks. | For, sublattice melting is known to occur in YSZ, and has been observed in YSZ/STO superlattices[20]. Against, sublattice melting observations are above CIC operating temperature, and when CIC operating conditions are examined no melting is found in computation (Fig 2a) or experiment[52]. |
| c. Phase change | For any $i, j$, $p_{ij}^{YSZ}(r) \neq p_{ij}^{CIC}(r)$. | For, YSZ is known to have multiple phases[22]. Against, no phase change has been observed under CIC operating conditions (Figs 2). |
| II. Mechanistic hypotheses | Required Proof | Evidence for and against. |
| a. Disorder | $\Delta ln(D) \sim \Delta S^{act}/k_B$ | For, structure is observably disordered - (Fig 2) and experiment[52].. Against, experiments demonstrate a marked Arrhenius dependence[17,25-28,44-47] (Fig 3). |
| b. Mechanics | $\Delta E_h^{diff} = \sigma_h V^{diff}$ | For, CIC effects are known to trends with superlattice structural misfit[17,19] (Fig 3). Against, magnitude of stress effect is not large enough and observed CIC demonstrates an orientation or texture[44,49] (Table 2, manuscript text). |
| c. Elastic Dipole | $\Delta E_{dip}^{mig} = \Delta \boldsymbol{P}^{act} \cdot \boldsymbol{\varepsilon}$ | For, CIC effects are known to trends with superlattice structural misfit[17,19], and the elastic dipoles have the required orientation effect. Against, magnitude of dipole effect is not enough (Table 2, manuscript text). |
| d. Space Charge | $\Delta E_{pot}^{diff} = q_{TS}\Delta\varphi_{TS} - q_R\Delta\varphi_R$ | For, space charge effects a known phenomenon in oxide interfaces, fits into existing phenomenology. Against, observed CIC demonstrates an orientation or texture[44,49]. |

**Table 4**. Summary of physics and evidence required for most commonly offered hypotheses controlling CIC.

**Supplementary Information**

**Structural hypotheses, limits on current conclusions.**

**The phase change hypothesis, more detailed required.** One of the posited hypotheses has been the possibility of a YSZ phase change inside of the CIC super-lattice environment. Here, this phase change hypothesis has been evaluated using an AIMD of 5 ps. This duration of 5 ps is adequate to evaluate several phenomena, such as the centrally important diffusion, however may not be sufficient to evaluate a phase change. In practice, there is no way to *a priori* know how long is required to simulate an unknown phase change. The intrinsic limitations of using AIMD to sample a phase change suggests the need for a thermodynamic analysis.

As the composition of $Y_xZr_yO$ changes, several phases are possible, where the stability of each of these phases may be influenced my strain, and by the presence of the interface. Predith *et al* have used a DFT and coupled cluster to examine the ordered ground states of $Y_xZr_yO$[22]. Over the span of the entire $Y_xZr_yO$ composition space, Predith *et al* have found 453 possible structures. When a thermodynamic analysis of these 453 structures was performed, the experimentally observed crystal structures was predicted to be, along with a new as yet unreported phase. To be sure, 453 is a large number of possible structurally variants, the energy of each of which might be affected by a strain. The stability of all of these phases as a function strain is not necessary to investigate, only those in the composition window relevant to CIC. The stability of these strained phases might consequently be influenced by the interfacial energy in the CIC heterolayer superlattice environment. To be sure, a complete analysis of the YSZ structure inside the CIC environment is a challenging, and as yet an incomplete task. Though the evidence here suggests cubic YSZ does not phase change, the large number of possible thermodynamic configurations is a cautionary note, and indicates that this particular conclusion is yet uncertain. To verify the possibility of phase change, either a detailed thermodynamic analysis must be performed, or *in situ* XRD.

Through the intrinsic time scale limitations of AIMD as a method of assessing phase stability draw scrutiny, it is important to remember that AIMD simulations of $cZrO_2$ have reproduced CIC. The reproduction of CIC within $cZrO_2$ means that while the heterolayer environment may later be shown to generate a phase change, it is unclear if this is important in the study of CIC.

**Mechanistic hypotheses, alternatives.**

Here have been tested the most frequently posed mechanistic hypotheses, and a tensorial version of one of those hypotheses, however, there are a few mechanistic hypotheses that have been posed, and deserve mention. These hypotheses are correlation[53], percolation[54], and string formation[55-58]. In addition to these untested hypotheses, a new one is here offered – fluctuations.

**Correlation**[53]. The correlation hypothesis poses that the oxygen vacancies do not migrate with a single hopping mechanism but rather in a collaborative motion with the impurity yttrium. Correlated comigration is a well-studied and well-observed phenomenon in metals[59-67], and therefore not without precedent. These studies of correlated diffusion in metals demonstrate that correlation effects can both facilitate and inhibit diffusion. Models for correlated diffusion are established for elemental metal crystal structures[62,63,66], though to the knowledge of this author, not for the cubic zirconia structure. In order to thoroughly test the correlation hypothesis, a model of correlated diffusion in cubic zirconia could be constructed, parameterized by DFT, and tested against AIMD simulations. The correlation model relies on this correlation process dominating diffusion, where here it has been shown that the activation barrier to hopping of the charged vacancy defect in $cZrO_2$ matches experimental measurements of the activation barrier in YSZ. Whether this evidence of the *operando* YSZ hopping mechanism is evidence against correlation, correlation has the great virtue of being testable.

**Percolation**[54]. The percolation hypothesis poses that the vacancy network has become sufficiently dense that continuous and facile vacancy diffusion pathways exist within the solid. AIMD simulations of the $cZrO_2$ CIC phenomenon have carefully added precisely one vacancy to the AIMD simulation[16,19,20]. The presence of only a single vacancy, and nonetheless the observation of CIC, may be taken as evidence against percolation, however, this conclusion too is incomplete. The interface between the YSZ and STO has not yet been well studied, and it may well be that under CIC operating conditions, both experimental and AIMD, the YSZ/STO interface stabilizes vacancies. These YSZ/STO interfacial vacancies may be sufficient to activate percolation. This hypothesis, too, has the virtue of being testable.

**String formation**[55-58]. The string formation hypothesis poses that diffusion events do not follow a thermodynamically determined Poisson process but are correlated in time. String formation is a well observed phenomenon in the diffusion of glasses and super-cooled liquids[68-71]. Though the $cZrO_2$ material responsible for CIC is a solid, and the oxygen sublattice is not melted, it is known to melt at higher temperatures[16,20], making the analogy with a super-cooled liquid applicable. This string diffusion mechanism is not mechanistically understood, but methods of separating string diffusion from thermal hopping diffusion have been developed[55-58]. This hypothesis, too, is testable.

**Fluctuations**[72,73]. Here is posed an additional hypothesis based on the correlation of fluctuations at the interface with the activation barrier inside of the electrolyte. In the testing of the charge hypothesis, and of possible elastic nonlinearity, the migration barrier to hopping within the interfacial environment was directly calculated and observed to change as a function of position, even for symmetric planes within the $cZrO_2$ material. The variation of the oxygen migration barrier within $cZrO_2$ is attributed to the inhomogeneity in the STO/YSZ interface. This interfacial inhomogeneity emits an inhomogeneous strain field which in turn affects the migration barriers. The variation in migration barrier may result in the hopping process becoming an extremal event process, controlled by a local activation barrier made possible by the strain field emanating from the interface. Precisely this type of fluctuation effect controls solute induced dislocation drag[72,73], twin migration in alloys[74], and the strength of high entropy alloys[75]. Too is this hypothesis testable.